\documentclass{mem}
\usepackage{natbib}\usepackage{txfonts}\usepackage{balance}
\usepackage{graphicx}
\usepackage{wrapfig}
\usepackage{multicol}
\usepackage{lipsum}
\usepackage{tikz}
\usepackage{float}
\usepackage[section]{placeins}
\usepackage{stfloats}
\usepackage[unskipbreak]{cuted}
\usepackage[font=small,labelfont=bf]{caption}
\usepackage{txfonts}
\usepackage{indentfirst}
\usepackage[utf8x]{inputenc}

\input{insbox}

\idline{-}{0}
\begin{document}
\def\teff{$T\rm_{eff }$}
\def\kms{$\mathrm {km s}^{-1}$}

\title{
Searches for TeV gamma-ray counterparts to Gravitational Wave events with H.E.S.S.
}

   \subtitle{}

\author{
H. \,Ashkar\inst{1} 
\and F. \,Schüssler\inst{1}
\and M. \,Seglar-Arroyo\inst{1} \\
(on behalf of the H.E.S.S. collaboration)
          }

\institute{
$^1$IRFU, CEA, Université Paris-Saclay, F-91191 Gif-sur-Yvette, France, \\
\email{halim.ashkar@cea.fr}
}

\authorrunning{H. Ashkar}

\titlerunning{INTEGRAL looks AHEAD to Multi-Messenger Astrophysics}

\abstract{
\indent The search for electromagnetic counterparts for gravitational waves events is one of the main topics of multi-messenger Astrophysics. Among these searches is the one for high energy gamma-ray emission with the H.E.S.S. Imaging Atmospheric Cherenkov Telescopes in Namibia. During their second Observation Run O2, the Advanced Virgo detector in Italy and the two advanced LIGO detectors in Washington and Louisiana while conducting joint observations, detected for the first time, on August $14^{th}$, 2017 a transient GW signal due to the coalescence of two stellar masses black holes, an event labeled GW170814. The alert announcing the event was issued two hours later and H.E.S.S. observations could be scheduled for the nights of $16^{th}$, $17^{th}$ and $18^{th}$ August 2017. Three days after the binary BH merger, on August $17^{th}$, the coalescence of two neutron star was detected for the first time, followed by a GRB detection by Fermi's GBM starting a new era in multi-messenger Astronomy. Observations started 5.3 h after the merge and contained the counterpart SSS17a that was identified several hours later. It stands as the first data obtained by a ground-based pointing instrument on this object. In this contribution, we will present the results of the search of high-energy gamma ray emission as electromagnetic counterpart of these two GW events. No significant gamma ray emission was detected for either event. Nevertheless upper limit maps were derived constraining, for the first time, the non-thermal, high-energy emission on the remnant of a three detector binary black hole coalescence (GW170814), and a binary neutron star coalescence (GW170817).
}
\maketitle{}

\section{Introduction}
\indent Since the observation of the Gravitational Waves (GW) event GW150914 emanating from the inward spiral and merger of two stellar masses black holes \citep{GW150914}, a new era in astronomy began. GWs have become the newest astronomical messenger joining electromagnetic waves, cosmic rays and neutrinos.\\
\indent The H.E.S.S. Imaging Atmospheric Cherenkov Telescope array is composed of one 28-m and four 12-m telescopes with field of views of 3.2$^{\circ}$ and 5$^{\circ}$. It is sensitive to gamma rays in the range of 50 GeV to 100 TeV, and is capable of detecting a point source at a 5 $\sigma$ level with a similar strength to the Crab in less than one minute \citep{H.E.S.S.sensitivity}. The preparation for the GW alerts from the advanced configurations of the LIGO and VIRGO interferometers within the H.E.S.S. collaboration started in summer 2016. After several technical and commissioning runs used to optimize the response to GW alerts, the first observation campaigns were conducted in August 2017.\\
\indent On August $1^{st}$ 2017 the Advanced Virgo interferometer joined the two LIGO interferometers in their second Observation Run O2. On August $14^{th}$ 2017, a GW signal was detected by the interferometers at 10h30m43s UTC. This signal was produced by the merger of two stellar masses black holes at a distance of $540^{+130}_{-210}$ Mpc, which corresponds to a redshift of z=$0.11^{+0.03}_{-0.04}$. The initial masses of the black holes were $30.5^{+5.7}_{-3.0}$ \(\textup{M}_\odot\) and $25.3^{+2.8}_{-4.2}$ \(\textup{M}_\odot\) \citep{GW170814}. This was the first detection by the three observatories and the added independent baselines from Virgo reduced the localisation uncertainty significantly. Three days later, on August 17, the coalescence of two neutron stars was detected for the first time \citep{GW170817} followed by a GRB detection by Fermi's GBM \citep{GRB}. Information from the three interferometers was used to compute a localisation area of ~30 deg$^2$.\\
\indent We here present for the first time the analysis of the H.E.S.S. data obtained during follow-up observations of GW170814. We also briefly summarize the results obtained during the prompt H.E.S.S. campaign on GW170817.

\section{GW170814}

\begin{figure}[h]
\begin{center}
\includegraphics[width=65mm]{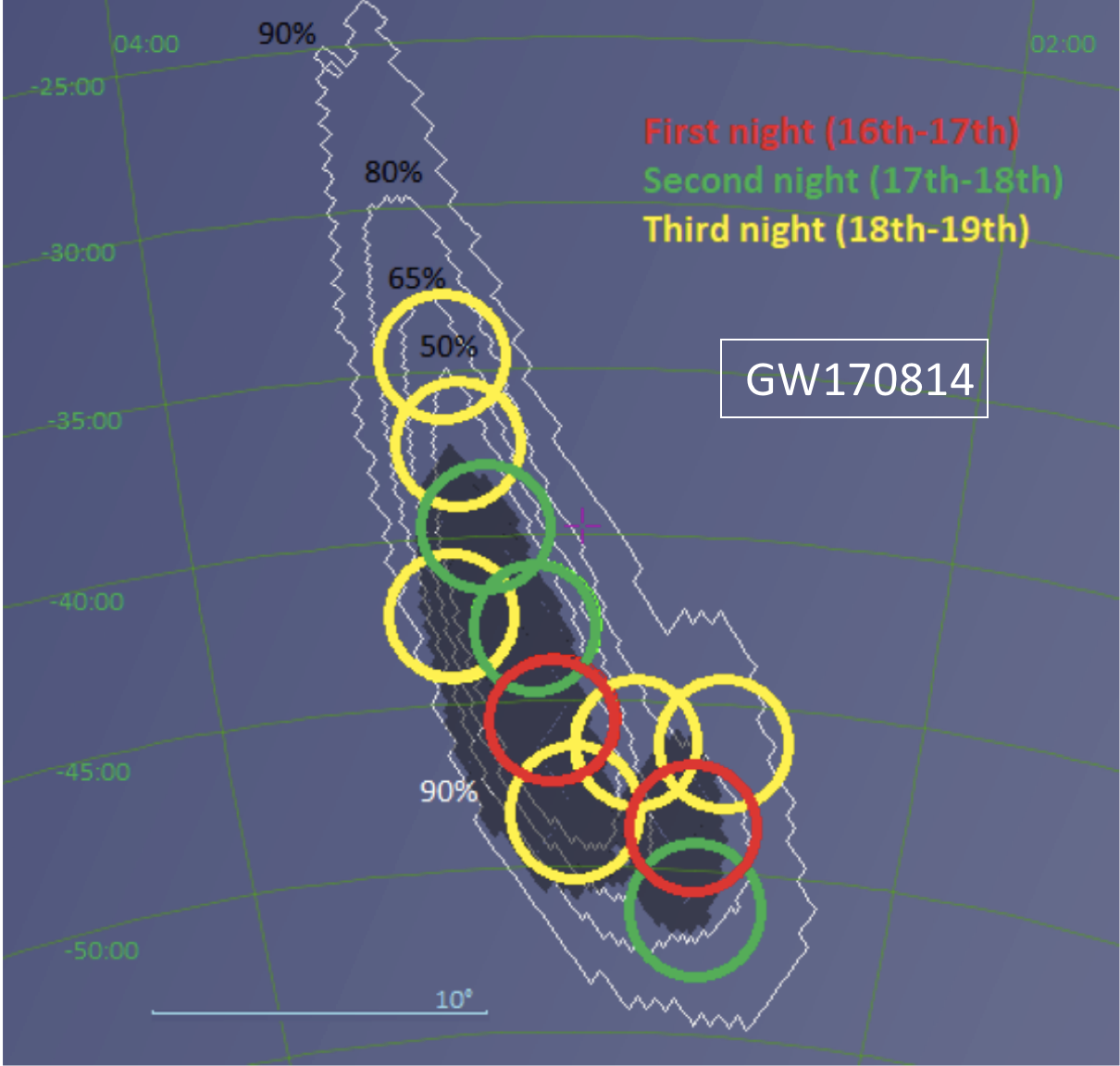}
\end{center}
\caption{\footnotesize
H.E.S.S. follow-up observations of GW170814. The white contours are the probability contours from the initial LALInference probability map which was used for the H.E.S.S. observation scheduling. The filled contour represents 90\% of the final LALInference map. The rings indicate the field of view of the H.E.S.S. observation runs throughout the three nights.}
\label{H.E.S.S.Obs}
\end{figure}

When reconstructing the localization of the GW emitter, there is a trade-off between latency and accuracy, which is represented by two different algorithms: the BAYESTAR algorithm \citep{BAYESTAR} is for rapid localisation and LALInference \citep{LALInference}, which is scanning a larger parameter space and marginalizing over calibration uncertainties is for a more precise but slower approach. The sky map used to schedule the H.E.S.S. observations was a LALInference based map and was issued on August $15^{th}$, 2017 at 20h17m51s UTC. It had an uncertainty region of 190 deg$^2$ due to the use of LIGO and Virgo data. However, after noise removal and offline detector calibration, a full parameter estimation constrained the 90\% credible region to a slightly shifted area of 60 deg$^2$. This is the final probability region that was published several weeks later \citep{GW170814}. In Fig. \ref{H.E.S.S.Obs} we show the different probability contours (in white) for the map used for the H.E.S.S. scheduling. Quoted values correspond to the \% containment probability region for the GW event localisation. The filled contour encloses a region of the updated (final) LALInference map with a 90\% probability of containing the event. We note that even thought the observation runs obtained by H.E.S.S. where scheduled using the initial LALInference map, it covers most of the final 90\% uncertainty region and the systematic shifts in the GW data don’t affect the H.E.S.S. capabilities to cover the region.\\
\indent During the GW170814 follow-up, the 28-m H.E.S.S. telescope was used alongside three 12-m telescopes. Observations started on August $17^{th}$ at 00h10m UTC and a total of 11 runs of 28 minutes each were obtained during three consecutive nights covering approximately 90\% of the final localisation region.

\begin{figure}[h]
\begin{center}
\includegraphics[width=65mm]{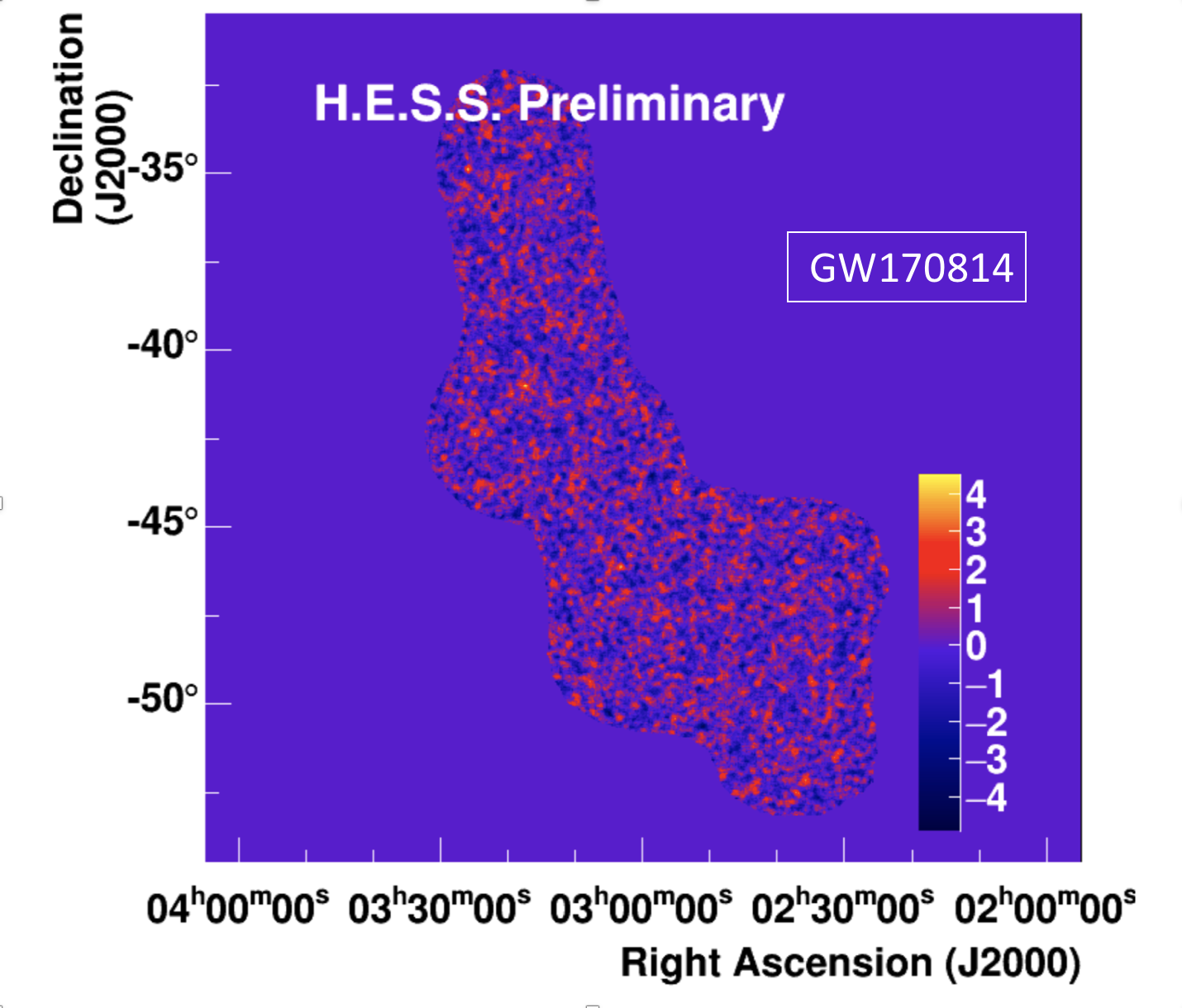}
\end{center}
\caption{\footnotesize VHE gamma-ray significance map in $\sigma$ units obtained from H.E.S.S. observations of GW170814.}
\label{H.E.S.S._SIGNIFICANCE}
\end{figure}

\begin{figure}[h]
\begin{center}
\includegraphics[width=65mm]{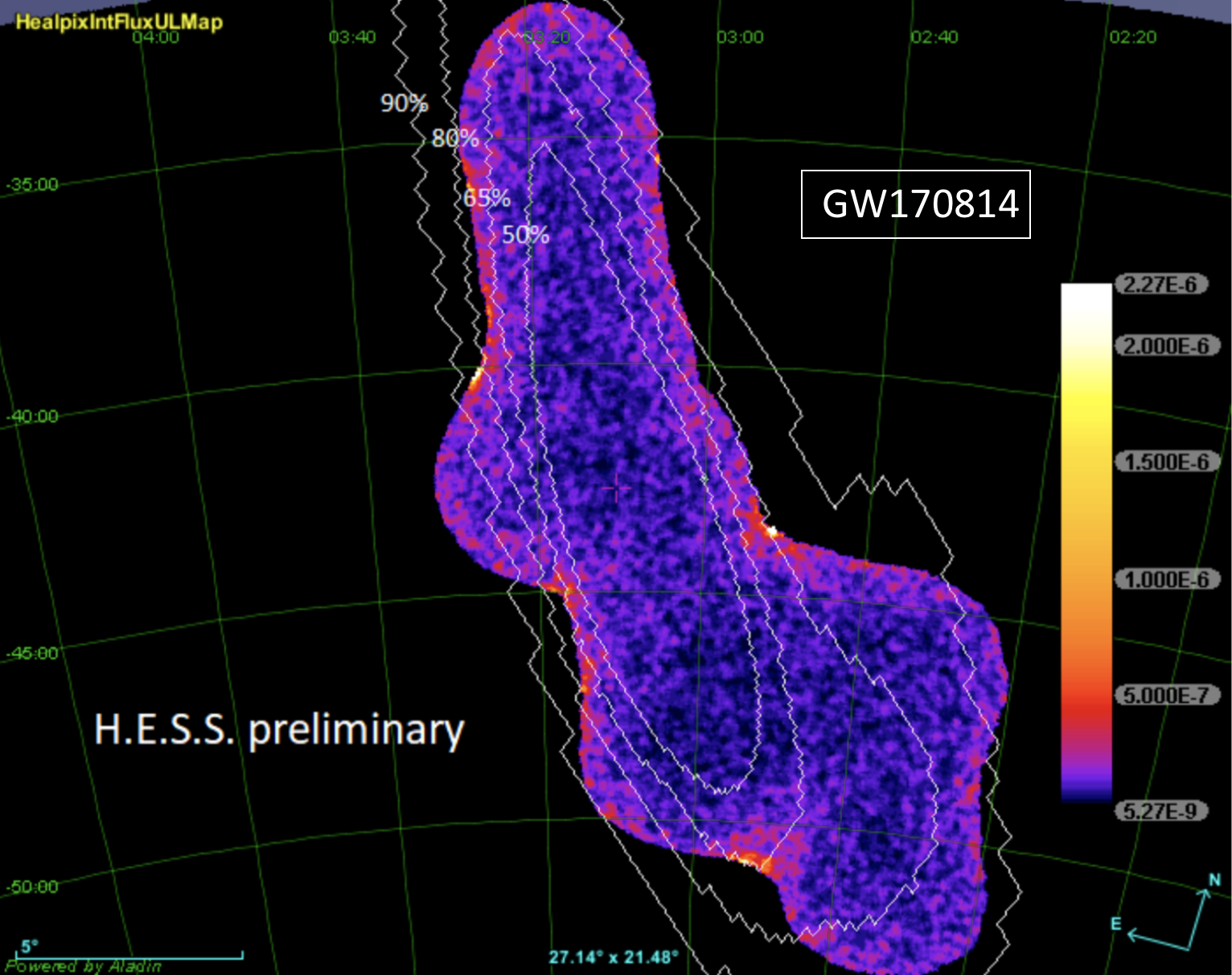}
\end{center}
\caption{\footnotesize Integral upper limits map in erg cm$^{-2}$ s$^{-1}$ derived from H.E.S.S. follow-up of GW170814 valid in the energy range 250 GeV $<$ E $<$ 20 TeV. The white contours are the probability contours from the initial LALInference probability map.}
\label{H.E.S.S._UL}
\end{figure}

\indent Fig. \ref{H.E.S.S._SIGNIFICANCE} shows the gamma-ray significance map which results from the analysis of the data obtained following the pointing pattern presented in Fig \ref{H.E.S.S.Obs}. No significant gamma-ray emission is found during this follow-up campaign.

\indent Combining all observations obtained with H.E.S.S. during the follow-up campaign of GW170814, we derive in Fig. \ref{H.E.S.S._UL} a sky map showing the integral upper limits to constrain the emission in the region for 250 GeV $<$ E $<$ 20 TeV assuming a generic $E^{-2}$ spectrum. Induced by the radially decreasing acceptance of the telescope, the obtained limits are less constraining when approaching the border of the field-of-view. This map can be used to constrain the emission of known gamma-ray sources in our region of interest but more importantly it allows for the first time a constraint to be placed on the level of very-high energy emission from the merger of two stellar mass black holes detected by both LIGO and Virgo detectors.

\section{GW170817}
\indent Extensive details on the GW170817 follow-up with H.E.S.S. can be found in \citep{H.E.S.S._GW170817}. Observations started 5.3 hours after the NS merger. The H.E.S.S. semi-automatic reaction and the implemented observation strategies allowed to react promptly and get on target 5 minutes after the release of the LALInference sky localisation. This allowed to get the first observations on the NS merger from a ground based observatory. Although no significant detection was reported, it allowed to place for the first time stringent upper limits for non-thermal emission for NS-NS mergers in the VHE gamma-ray domain (270 GeV $<$ E $<$ 8.55 TeV). Fig. \ref{H.E.S.S._GW170817} summarizes the follow-up campaign of GW170817 with H.E.S.S.

\begin{strip}
\InsertBoxC{
\includegraphics[scale =0.40]{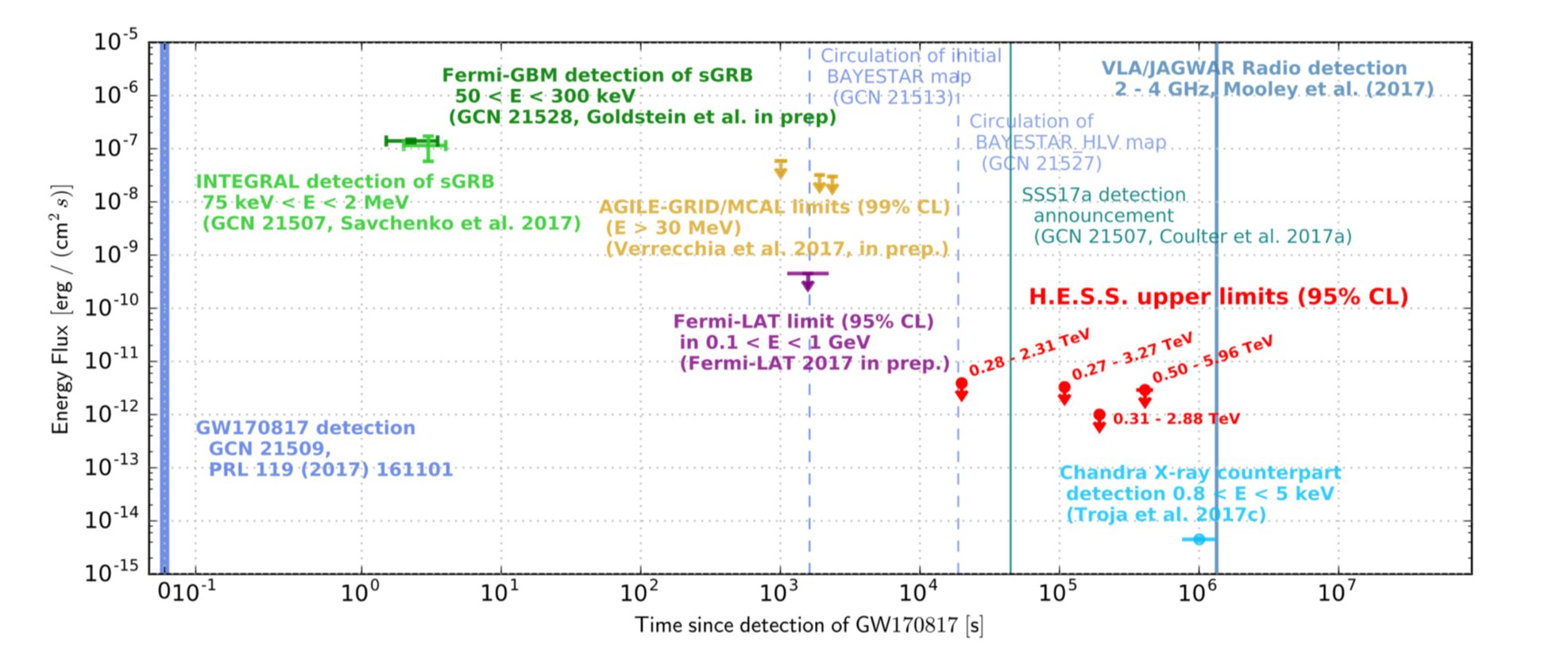}}
\captionof{figure}{Timeline of the observations following the detection of GW170817 with a focus on the high-energy, non-thermal domain.}
\label{H.E.S.S._GW170817}
\end{strip}

\section{Conclusion}
\indent In this contribution we summarized the most interesting H.E.S.S. observations of GW events during O2. Even though no significant VHE gamma-ray emission was detected, we are able to constrain the non-thermal emission from these events by computing integral upper limits on the non-thermal emission of the remnants. These successful observation campaigns provided also important feedback for the next observation run O3. With the approach of O3, new follow-up strategies that further increase the speed of the H.E.S.S. reaction to GW events have been fully implemented and tested within the H.E.S.S. experiment and are used to prepare the multi-messenger program of the upcoming future Cherenkov Telescope Array (CTA).

\bibliographystyle{aa}

\end{document}